\documentclass[12pt]{article}
\usepackage[english]{babel}
\usepackage{graphics} 
\usepackage{psfig}
\newcommand{\beq}{\begin{equation}}
\newcommand{\eeq}{\end{equation}}
\newtheorem{theorem}{Theorem}
\newtheorem{proof}{Proof}

\begin{document}
\Large
\centerline{\bf Scalar solitons in a 4-Dimensional curved space-time}
\large
\vskip 0.8cm
\centerline{Jos\'e A. Gonz\'alez\footnote{cervera@nuclecu.unam.mx} and 
Daniel Sudarsky\footnote{sudarsky@nuclecu.unam.mx}}

\normalsize
\centerline{\it Instituto de Ciencias Nucleares}
\centerline{\it Universidad Nacional Aut\'onoma de M\'exico}
\centerline{\it A.P. 70-543, M\'exico, D.F. 04510, M\'exico.}

\abstract{There is a theorem known as a Virial theorem 
that restricts the possible existence of 
non-trivial static solitary waves with scalar fields in a flat space-time with 
3 or more spatial dimensions. This raises the following
question: Does the analogous curved space-time version hold?. We investigate 
the possibility of solitons in a 4-D curved space-time with a simple model 
using numerical analysis. We found that there exists a static solution of the
proposed non linear wave equation. This proves that in curved space-time the 
possibilities of solitonic solutions is enhanced relative to the flat 
space-time case. }
\vskip 0.6cm

\section{Introduction}
Solitons are special solutions of non-linear wave equations. The most 
relevant characteristic of solitons is that they 
are localized static solutions. The simplest example consist of a single 
scalar field $\phi$ in one spatial and one temporal dimensions. 
Perhaps the most famous one is the sine-Gordon soliton \cite{1},\cite{2}.

At first sight, it can be thought that a wave equation with a single scalar 
field in more than three spatial dimensions with solitonic solutions can be 
found. However there is a Virial theorem which restricts this possibility.
Here we are going to transcript that theorem and it's proof for the 
convenience of the reader \cite{3}:

\begin{theorem}
There are no non-trivial static solitary waves of systems with scalar fields 
when the space dimensionality is three or more and when the Lagrangian has 
the form:
\beq
\cal L(\bf x\rm,t)=\frac{1}{2}(\partial_\mu \phi)(\partial^{\mu} \phi)-U(\phi(\bf x\rm,t))
\eeq
with $\phi=[\phi_i(\bf x\rm,t);i=1,...,N]$ a set of N coupled scalar fields in 
D space plus one time dimensions, and $U(\phi(\bf x\rm,t))$ a positive 
definite potential.
\end{theorem}

\begin{proof}
A static solution $\phi(\bf{x})$ obeys
\beq
\nabla^2 \phi=\frac{\partial U}{\partial \phi}(\bf{x})
\eeq
where $\nabla^2$ is the Laplacian in D dimensions. This equation clearly 
correspond to the extremum condition $\delta W=0$ for the static energy functional
\beq
W[\phi]\equiv \int d^D x \left[ \frac{1}{2} \nabla_i \phi \cdot \nabla_i \phi + U(\phi(\bf{x})) \right] \equiv V_1[\phi]+V_2[\phi]
\eeq
where the functionals $V_1$ and $V_2$ stand for the two terms on the right-hand 
side. Note that not only W but also $V_1$ and $V_2$ are non-negative. 
Now, let $\phi_1(\bf{x})$ be a static solution. Consider the one-parameter 
family of configurations
\beq
\phi_\lambda=\phi_1(\lambda\bf{x}).
\eeq
It is easy to check that
\beq\label{5}
W[\phi_\lambda]=V_1[\phi_\lambda]+V_2[\phi_\lambda]=\lambda^{2-D}V_1[\phi_1] + \lambda^{-D}V_2[\phi_1].
\eeq
Since $\phi_1(\bf{x})$ is an extremum of $W[\phi]$, it must in particular 
make $W[\phi_\lambda]$ satationary with respect to variations in $\lambda$; 
that is,
\beq\label{6}
\frac{d}{d\lambda}W[\phi_\lambda]=0 \hskip 1cm at \hskip 0.4cm \lambda=1.
\eeq 
Differentiating (\ref{5}) using (\ref{6}) gives us
\beq\label{7}
(2-D)V_1[\phi_1]=DV_2[\phi_1].
\eeq 
Since $V_1$ and $V_2$ are non-negative (\ref{7}) cannot be satisfied for 
$D\ge3$ unless $V_1[\phi_1]=V_2[\phi_1]=0$.
This means that $\phi_1(\bf{x})$ has to be space-independent and equal to 
one of the zeros of $U[\phi]$. This is just a trivial solution and the 
theorem precludes non-trivial space-dependent solutions. \bf{q.e.d.}
\end{proof} 

For the case of $D=2$, equation (\ref{7}) tell us that $V_2[\phi_1]=0$. The 
simplest example of a solution of this kind is the non-linear O(3) model 
\cite{4},\cite{5},\cite{6} relevant in the description of the statical mechanics of an 
isotropic ferromagnet.

We are interested in the possible existence of a static solution of a system 
in a curved space-time in three spatial and one temporal dimensions. We will 
construct a simple model and look for solitonic solutions numerically.

\section{The model} 

Consider the simplest case in a curved space-time. One scalar field whose 
equation of motion is:
\beq\label{8}
\partial^{\mu}\partial_{\mu}\phi-\frac{\partial V}{\partial \phi}=0
\eeq
Without potencial (\ref{8}) is the Laplace equation,whose solutions take a 
maximum or minimum value only at the spatial boundaries. If we solve the 
Laplace equation in a space with a connected boundary or in a compact space, 
the solution in every point of the space will necessarily have the same 
value as in the boundary (i.e. a trivial solution).

So we are going to introduce a potential $V=(\phi^2-1)^2$ in analogy with the 
$1+1$ dimensional case. Moreover, we consider the simplest kind of spatialy 
compact universe, the static Einstein Universe:
\beq\label{9}
ds^2=-dt^2+d\chi^2+sin^2\chi(d\theta^2 + sin^2\theta d\varphi^2).
\eeq
Equation (\ref{8}) in the space-time corresponding to (\ref{9}) takes the form:
\beq
\frac{\partial^2\phi}{\partial\chi^2}-\frac{\partial^2\phi}{\partial t^2}
+2cot\chi\frac{\partial\phi}{\partial\chi}
+sin^2\chi cot\theta\frac{\partial\phi}{\partial\theta}
+sin^2\chi\frac{\partial^2\phi}{\partial\theta^2}
+sin^2\chi sin^2\theta\frac{\partial^2\phi}{\partial\varphi^2}=2(\phi^2-1)2\phi.
\eeq

We are interested in static solutions and with spherical simetry so we have
\beq\label{11}
\frac{d^2\phi}{d\chi^2}+2cot\chi\frac{d\phi}{d\chi}-4\phi(\phi^2-1)=0
\eeq
this is an ordinary differential equation of second order. Equation (\ref{11}) 
can be separated in 2 ordinary equations of first order. Let:
\beq
x_1=\phi \hskip 1cm  x_2=\frac{d\phi}{d\chi}
\eeq
so, (\ref{11}) can be written as the following system:
\beq
x_2=\frac{dx_1}{d\chi}
\eeq
\beq
\frac{dx_2}{d\chi}+2x_2cot\chi-4x_1(x^2_1-1)=0.
\eeq

These equations can be integrated by the Runge-Kutta method \cite{7}.
\vskip 0.5cm

We note that equation (\ref{11}) is singular at $\chi=0$ and $\chi=\pi$, 
so we use the following initial conditions:
\beq
\phi(0)=\phi^0_a \hskip 1cm \frac{d\phi(0)}{d\chi}=\dot{\phi}^0_a=0
\eeq
\beq
\phi(\pi)=\phi^0_b \hskip 1cm \frac{d\phi(\pi)}{d\chi}=\dot{\phi}^0_b=0
\eeq
So we have a differential equation (\ref{11}) with boundary conditions at the 
two end points.
It can be solved using a ``shooting to a middle point'' method, with 
$\phi^0_a$ and $\phi^0_b$ as shooting parameters. The main idea of this 
method is the following:

We perform an integration from $\chi=0$ to $\chi=\frac{\pi}{2}$ obtaining 
$\phi_a(\frac{\pi}{2})$. Also we integrate from $\chi=\pi$ to 
$\chi=\frac{\pi}{2}$ obtaining $\phi_b(\frac{\pi}{2})$.
Now, let construct the following function:
\beq
F(\phi_a,\phi_b)=\left[\phi_a\left(\frac{\pi}{2}\right)-\phi_b\left(\frac{\pi}{2}\right),\dot{\phi}_a\left(\frac{\pi}{2}\right)-\dot{\phi}_b\left(\frac{\pi}{2}\right)\right]
\eeq

The first integration allows us to evaluate F obtaining in general 
$F(\phi^0_a,\phi^0_b)\neq 0$ (figure 1). We are interested in the zeros of 
this function, because at those points the solutions of the integration 
match $\phi$ and $\dot{\phi}$ in a smooth way. We used the Newton-Raphson 
method to find the zeros of a function \cite{7}, and the result is showed
in figure 2.

\begin{figure}[htp]
\centerline{
\psfig{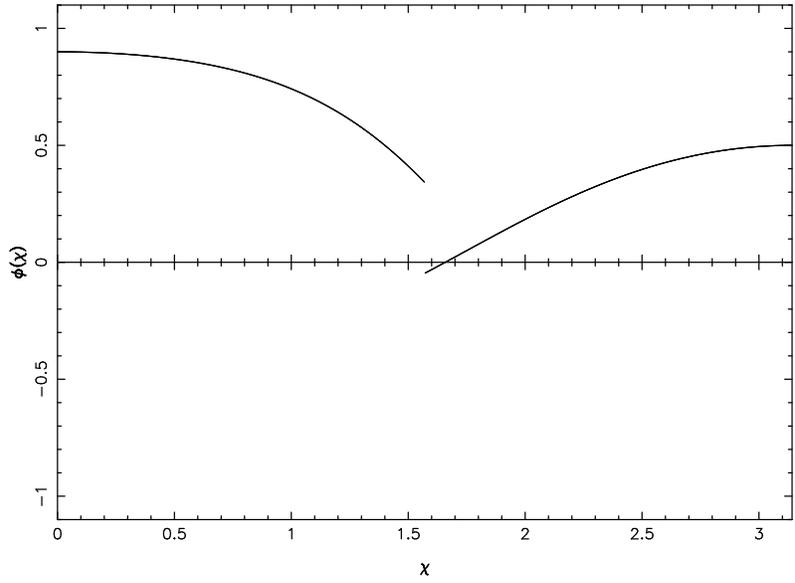}}
\caption{\small Integration for the initial values $\phi^0_a$ and $\phi^0_b$. 
Obviously $\phi_a(\frac{\pi}{2})\neq\phi_b(\frac{\pi}{2})$ and $\dot{\phi}_a(\frac{\pi}{2})\neq\dot{\phi}_b(\frac{\pi}{2})$.}
\label{graf1}
\end{figure} 

\begin{figure}[htp]
\centerline{
\psfig{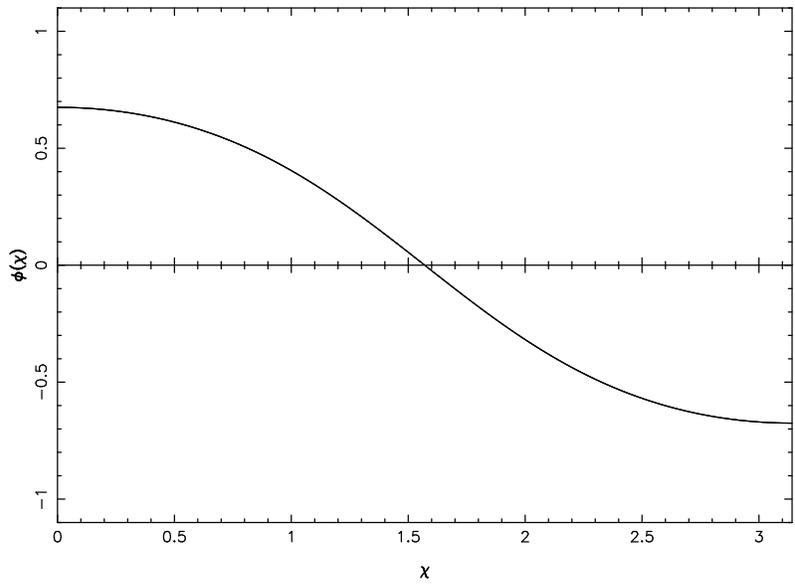}}
\caption{\small Integration after the application of the Newton-Raphson method. 
We can see that $\phi_a(\frac{\pi}{2})=\phi_b(\frac{\pi}{2})$ and $\dot{\phi}_a(\frac{\pi}{2})=\dot{\phi}_b(\frac{\pi}{2})$. 
This is a solution of the equation of motion (11).}
\label{graf2}
\end{figure}

\section{Conclusion}

We have shown that the theorem that precludes the existence of solitonic 
solutions to systems based in saclar field in $3+1$ or more flat space-time 
dimensions would be false if extended to the curve space-time case. 
We have done so by explicitly constructing one such solitonic solution of a 
simple model based on a single scalar field in the static Einstein Universe.

This is another indication\footnote{The most widely knows example of this 
phenomenon arose in the consideration of Einstein-Yang-Mills theory where 
solitonic solutions have been found while it know that these are no such 
solutions in Yang-Mills theory in Minkowskian space-time \cite{8}.} that the 
interplay of curved space-time physics and soliton physics allow us for
richer phenomena than each either of the two fields on its own.

\section{Aknowledgments}
This work was partially supported by DGAPA-UNAM grant 
\#IN121298 and CONACyT grant \#32272E.

\end{document}